\documentclass[aps,12pt,showpacs,showkeys]{revtex4-1}
\usepackage{graphics,epsfig,subfigure}
\usepackage{amssymb,amsfonts}
\usepackage{color,psfrag}
\usepackage{pgf,tikz}

\newcommand{\nn}{\nonumber}
\newcommand{\bb}[0]{\begin{eqnarray}}
\newcommand{\ee}[0]{\end{eqnarray}}

\newcommand{\eref}[1]{Eq.~(\ref{#1})}
\newcommand{\efig}[1]{Fig.~\ref{#1}}
\renewcommand{\Re}{{\rm Re}}

\newcommand{\Ar}{A_0}
\newcommand{\ff}{\makebox{\small $\frac{1}{2}$}}

\newcommand{\om}{\omega_c}
\newcommand{\tk}{\tilde\kappa}

\newcommand{\erf}{{\rm erf}}

\renewcommand{\d}{{\rm d}}
\newcommand{\freq}{{\delta F}}

\begin{document}
\title{Effect of electronic band dispersion curvature on de Haas-van Alphen oscillations}
\date{\today}
\author{Jean-Yves Fortin}
\email[E-mail address: ]{fortin@ijl.nancy-universite.fr}
\affiliation
{Institut Jean Lamour, D\'epartement de Physique de la Mati\`ere et des Mat\'eriaux,
Groupe de Physique Statistique, CNRS - Nancy-Universit\'e
BP 70239 F-54506 Vandoeuvre les Nancy Cedex, France\\
}
\author{Alain~Audouard}
\email[E-mail address: ]{alain.audouard@lncmi.cnrs.fr}
\affiliation{Laboratoire National des Champs Magn\'{e}tiques
Intenses (UPR 3228 CNRS, INSA, UJF, UPS) 143 avenue de Rangueil,
F-31400 Toulouse, France.}

\begin{abstract}
The effect of electronic band curvature, i.e. the deviation from parabolicity of
electronic dispersion, on de Haas-van Alphen oscillations spectra is studied.
Although the oscillations amplitude remain unaffected, it is demonstrated that
non-quadratic terms of the Landau bands dispersion, which is particularly relevant in the case of Dirac fermions, induces a field- and
temperature-dependent Onsager phase. As a result, a temperature-dependent shift of the de
Haas-van Alphen oscillations frequency is predicted.
\end{abstract}

\maketitle

\section{Introduction}

Magnetic oscillations or de Haas-van Alphen effect (dHvA) in 
quasi-two-dimensional metals are 
well accounted for by the Lifshitz-Kosevich (LK) 
theory~\cite{LandauVol9,Abrikosov,Ziman,LK2bis,LK2}, which relates
the frequency to the surface area enclosed by the cyclotronic trajectories.
This geometrical approach is based on the semi-classical quantization theory of
Onsager~\cite{Onsag52}, and allows for the determination of many physical
parameters of the Fermi surface (FS). Characteristic field $B_c$ =
$\Phi_0$/$\Ar$  associated with the
quantum flux trough the unit cell area $\Ar$ is generally very large.
Indeed, for organic conductors with unit cell area as large as $\Ar$ =
100${\rm \AA}^2$, $B_c$ is still 4136 T. Therefore, available magnetic
fields stay  well within the limit of the semi-classical approximation. In the
opposite case, e.g. for significantly larger unit cell or applied magnetic
fields,  quantum corrections to the Landau
spectrum~\cite{rabe} or modification of the Lifshitz-Kosevich (LK) theory 
would be necessary.

A question that remains to be addressed deals with the effect of departure from parabolic curvature of the electronic band dispersion on the amplitude and phase of quantum oscillations. This question is particularly relevant in the case of Dirac fermions, the electronic dispersion of which is linear. In this case, the LK calculation
based on a truncation at first order in energy when evaluating the grand
potential is no
more sufficient, since the Landau level energy is known to display a square root
dependence on the Landau level index~\cite{mcclure66}.

In this paper, we address the
question of the relevance of the non-parabolicity in two classes of materials. 
First, the LK calculation for magnetization in the presence of a 
uniform field is reconsidered in the case of Landau quasiparticles, or
conventional fermions, relevant to
e.g. organic conductors. Then Dirac fermions, which have linear
band dispersion, are considered. In both cases the field- and
temperature-dependent phase factor of the Fourier coefficients is evaluated as 
a function of the FS curvature which is the main parameter of the model. We can 
also mention other work dealing with non-parabolicity of the Fermi surface, in 
the special case of a tight-binding model in two dimensions~\cite{Yong:1996} 
where the band gap closes for certain filling factor, and where the temperature 
amplitude does not follow the LK formula.

The main results of this paper can be summarized as follow. For a band 
dispersion in a two dimensional material with a closed Fermi surface, we assume 
that the area of such surface is given by the quantity $S(E)$ at energy $E$, 
close to the Fermi energy $E\simeq \mu$. 
The effect of band curvature (defined as the second 
derivative $S''(E)=\partial^2S(E)/\partial E^2$ near the Fermi energy $E=\mu$) 
is 
mainly to add a phase shift contribution $\phi_p$ 
in each harmonics of order 
$p$ of the oscillating quantities such as the 
magnetization. More precisely, the phase shift is a function of field and 
temperature and is given by the following semi-classical expression in the 
case of organic conductors
\bb\nn
\phi_p&=&\frac{\pi^2 k_BT\hbar^2S''(\mu)}{m^*}\varphi\Big (
2\pi^2 p \frac{k_BT}{\hbar\om}\Big ),
\\ \label{phi}
\varphi(x)&=&\Big [ 
\sinh(2x)-x-x\cosh^2(x) \Big ]/\sinh^2(x),\;\;
\ee

where $m^*$ is the effective mass of the quasi-particle and $\om=eB/m^*$ the 
cyclotronic frequency.
In the case of Dirac fermions, for which the Landau energy level spectrum $E_n$ 
increases like the square root of the index $n$, the expression of the phase 
shift is given instead by 
\bb\nn
\phi_p&=&\frac{\pi}{2}\frac{k_BT}{m^*v_F^2}\varphi\left (
2\pi^2p\frac{k_BT}{\hbar\omega_c} \right ),
\\ \label{phiGraph}
\varphi(x)&=&\Big
[ 2\sinh(2x)-x-x\cosh^2(x)-2x^{-1}\sinh^2(x) \Big ]/\sinh^2(x).
\ee
where $v_F$ is the Fermi velocity. In the following we will derive 
both~\eref{phi} and~\eref{phiGraph} using semi-classical analysis, and 
study their asymptotic limits, when both $T$ or $B$ are varied within physical 
ranges. This effect has to be differentiated from other possible contributions 
coming from additional physical mechanisms. For example, the presence of a 
spin-orbit coupling leads to a splitting of the dHvA frequency whose magnitude 
is proportional to $B^2$ and effective mass $m^*$~\cite{Mineev:2005}. Frequency
splitting due to spin-orbit coupling has been studied in bilayer high-T$_c$ 
cuprates~\cite{Sebastian:2014} and can also be attributed to the splitting of 
the Fermi surface. Additional phases exist in the presence of magnetic 
breakdown. It is well known that magnetic breakdown is accompanied 
with a field-dependent phase~\cite{Sl67}, especially for large orbits. This 
phase depends more precisely on the ratio between the field and the breakdown 
field, and not temperature. Existence of this Onsager phase has been studied 
and revealed in organic 
compound $\theta$-(BEDT-TTF)$_4$CoBr$_4$(C$_6$H$_4$Cl$_2$) where the breakdown 
field is close to 35T~\cite{2013SyntM}. 
More recently, the existence of phase offsets $\gamma$ has been 
questioned in thermodynamic 
quantities of three-dimensional topological insulators with surface states 
~\cite{Mikitik:1999,Lukyanchuk:2004,Wright:2013}. The topological 
nature of these insulators can be detected due to the presence of a Berry 
phase within the oscillations when the particle-hole symmetry is broken and 
the material has a band gap~\cite{Wright:2013b}. These global phases (as for 
the Maslov index $\gamma$, see below) are independent of temperature, 
but present a linear variation with the field~\cite{Wright:2013}, contrary to 
the field and temperature-dependent phases~\eref{phi} and~\eref{phiGraph} which 
come from a local effect of the band dispersion as we will see in the next 
section. Non-zero topological Berry's phase was also investigated in 
graphene~\cite{Zh05} by measuring the magnetoresistance in the quantum Hall 
regime, with a compelling evidence of a value for $\gamma$ different from 1/2 
due to the presence of Dirac fermions. For a review of the topological phases in
condensed matter physics, see~\cite{Senthil:2015}.

\section{Effect of band curvature corrections in organic conductors}

Within the semi-classical framework, the phase quantization can be expressed in terms of surface
swept by the quasiparticle in the Brillouin zone. It is given by
the integral
\bb\label{SEn}
S(E_n=E)&=&\frac{1}{4\pi^2}\oint_{E=E(k_x,k_y)}
k_y\,dk_x=\pm b(n+\gamma),
\ee
where $b = eB/h$ is an effective Planck constant or reduced field,
and $E_n$ is the energy of the Landau band with $n$ the Landau level index. 
$\gamma$ is the Maslov index which is
equal to $1/2$ for Landau quasiparticles with a parabolic band. In such a case,
$S(E)=m^*E/(2\pi\hbar^2)$ varies linearly with the energy. For massless Dirac electrons
with a linear energy
dispersion, $S(E)=\pi E^2/(2\pi\hbar v_F)^2$  and $\gamma = 0$.
In two-dimensional systems, the grand potential is expressed by
\bb\nn
\Omega(B,\mu,T)&=&-A\frac{b}{\beta}\sum_{n=0}^{\infty}
\log(1+\exp[\beta(\mu-E_n)]),
\ee
where $\mu$ is the  chemical potential and  $A$ is the sample area.
$bA$ is the degeneracy of each Landau Level.
The Poisson formula can be used for any discrete series over positive integers
$n$:
$\sum_{n=0}^{\infty}F(n)=\int_0^{\infty}F(n)\,\d n+2\sum_{p=1}^{\infty}
\int_0^ { \infty}F(n)\cos(2\pi pn)\,\d n$. This allows us to rewrite the
oscillating part of the grand potential in terms of Fourier modes
\bb\label{omega0}
\frac{\Omega_{osc}}{A}=-\frac{2b}{\beta} \Re
\sum_{p=1}^{\infty}
\int_0^{\infty}\log(1+\exp[\beta(\mu-E_n)])\exp(2i\pi pn)\,\d n.
\ee
Using $\exp(2i\pi pn)\rightarrow
\exp(2i\pi pn)/(2i\pi p)$ as primitive function, a double integration by parts 
can be
performed, yielding
\bb\nn
\hspace{-1.5cm}
\frac{\Omega_{osc}}{A}&=&-2b\,\Re
\sum_{p=1}^{\infty}
\int_0^{\infty}
\left [
\frac{\beta E'^2_n}{4\cosh^2[\frac{\beta}{2}(E_n-\mu)]}
\frac{\exp(2i\pi pn)}{(2i\pi p)^2} \right .
\\ \label{omega1}
&-&\left .
\frac{\beta E'_n E''_n}{4\cosh^2[\frac{\beta}{2}(E_n-\mu)]}
\frac{\exp(2i\pi pn)}{(2i\pi p)^3}
\right]\,\d n,
\ee
where $E'_n=\partial E_n/\partial n$ and $E''_n=\partial ^2E_n/\partial n^2$.
Since for a parabolic band $E''_n=0$,
only the first term does not vanish in this case and  $E'_n=\hbar\om$ is
$n$-independent and
proportional to $b$.
In general $E''_n, E'''_n$, etc... are non zero, and \eref{omega1}
can be solved around $E = \mu$ using the formal series expansion (see 
also~\cite{rabe})
\bb\label{Sa}                                 
S(E)=S(\mu)+(E-\mu)\frac{\partial S}{\partial E}(\mu)
+\ff (E-\mu)^2\frac{\partial^2 S}{\partial E^2}(\mu)+\cdots
\ee
In the standard LK theory, only the first two terms in~\eref{Sa} are taken into 
account: the first one typically sets 
the frequency of the oscillations, while the second one (linear in the energy 
difference) gives rise to the thermal reduction factor (see page 184 
of reference~\cite{Abrikosov} after equation 10.28 for 
instance). The third term studied in this manuscript, and quadratic in energy 
difference, is responsible for a phase shift of the oscillations as we will see 
further below.
To illustrate the discussion, we can take a typical example, the tight binding 
model of free electrons on a discrete lattice with hopping parameter $t$ and 
which is described by the energy dispersion
$E(k_x,k_y)=-t[\cos(k_x)+\cos(k_y)]$. The density of states
can be written as
\bb
\label{Eq5}
\Ar S(E)=\frac{4}{\pi^2}\int_{-2}^{E/t}\frac{1}{2-u}K\Big [
\left (\frac{2+u}{2-u} \right ) \Big ]\,\d u,
\ee
where $\Ar$ is the area of the unit cell and $K(k)$ is the
complete elliptic
integral of the first kind:
$K(k)=\int_0^1\,\d u [(1-u^2)(1-k^2u^2)]^{-\ff}$.~\eref{Eq5} is computed from 
\eref{SEn} using the coordinate equation $k_y=\cos^{-1}(-E/t-\cos(k_x))$, 
then differentiating it
with respect to the energy to get $S'(E)$. After the change of
variable $u=\cos(k_x)$,~\eref{Eq5} can be
rewritten as
\bb
\Ar \frac{\partial S(E)}{\partial E}=\frac{1}{\pi^2t}
\int_{-1-E/t}^1\frac{\d u}{\sqrt{(1-u^2)[1-(u+E/t)^2]}}.
\ee
After the additional change of variable $v=u+\ff E/t$, the integral is symmetric
around the origin, and a further transformation $w=v/(1+\ff E/t)$ leads to an
expression involving the elliptic integral. We can in particular perform an
expansion around the lower band limit $E=-2t$ such that
\bb
\Ar S(E)=\frac{1}{2\pi}\Big (\frac{E}{t}+2\Big )+\frac{1}{16\pi}\Big
(\frac{E}{t}+2\Big )^2+\cdots
\ee
We can identify $S'(E)=(2\pi \Ar t)^{-1}$ with $2\pi
m^*/\hbar^2$, since $4t=4\hbar^2/(\Ar m^*)$ is the bandwidth.
The curvature parameter of the surface area enclosed by the orbit defined in 
the following by $\kappa=S''(E)$, can be rewritten as 
$\kappa=
(8\pi\Ar t^2)^{-1}$. Therefore for a given dispersion, we can 
relate the different coefficients of the expansion in~\eref{Sa} with 
microscopic parameters such as the hopping constants, effective mass or 
bandwidths. 

In the most general case, we would like to use the 
expansion~\eref{Sa} up to second order to compute the different thermodynamic 
quantities such as~\eref{omega0}. Up to now, the LK calculation considers 
only the first order around the Fermi surface, whose coefficient is given by 
the effective mass (the slope of $S(E)$). The second order term will modify, as 
we will see below, mostly the phase of the magnetic oscillations.
To include the effect of the second order term, we perform first a change of 
variable $n\rightarrow E$ in the expression of the grand potential 
\eref{omega1} using 
$dE=E_n'dn$
\bb\nn
\frac{\Omega_{osc}}{A}&=&-2b\,\Re
\sum_{p=1}^{\infty}
\int_{0}^{\infty}
\left [
\frac{\beta E'_n}{4\cosh^2[\frac{\beta}{2}(E-\mu)]}
\frac{\exp[2i\pi pn(E)]}{(2i\pi p)^2}
\right .
\\ \nn
&-&\left .
\frac{\beta E''_n}{4\cosh^2[\frac{\beta}{2}(E-\mu)]}
\frac{\exp[2i\pi pn(E)]}{(2i\pi p)^3}
\right]\,\d E.
\ee
We then use the relations
\bb
E'_n=\frac{\hbar\om}{1+\tk(E_n-\mu)},\;
E''_n=-\frac{\tk(\hbar\om)^2}{[1+\tk(E_n-\mu)]^2},
\;\tk=\frac{2\pi\hbar^2\kappa}{m^*},
\ee
to obtain the oscillating part of the grand potential expressed as an integral
over the energy
\bb\nn                                     
\frac{\Omega_{osc}}{A}&=&-2b\,\Re
\sum_{p=1}^{\infty}
\int_{0}^{\infty}
\left [
\frac{\beta\hbar\om}{4[1+\tk(E-\mu)]
\cosh^2[\frac{\beta}{2}(E-\mu)]}
\frac{\exp[2i\pi pn(E)]}{(2i\pi p)^2}
\right .
\\ \label{omega2}
&+&
\left .
\frac{\tk\beta (\hbar\om)^2}{4[1+\tk(E-\mu)]^2\cosh^2[\frac{\beta}{2}(E-\mu)]}
\frac{\exp[2i\pi pn(E)]}{(2i\pi p)^3}
\right]\,\d E,
\ee
where $n(E)=S(E)/b-\gamma$, $S(E)$ being given by the expansion \eref{Sa}
around the Fermi energy.
The next step is to perform the integration around the saddle point $E=\mu$ at
low temperature, using the variable $x=\beta(E-\mu)/2\pi$ and replacing
$n(E)$ by $n(x)$
\bb                                                                             
n(x)=-\gamma+\frac{S(\mu)}{b}+\frac{2\pi
x}{\beta\hbar\om}+\frac{2\pi^2x^2\tk}{ \beta^2\hbar\om}.
\ee
In the case of a parabolic band, the first and second terms are related to the
Onsager phase or Maslov index, and the oscillation frequency
$F=hS(\mu)/e$, respectively. More specifically, magnetization is
obtained by differentiation of \eref{omega2} with respect to minus $B$, yielding
\bb\label{mosc}
m_{osc}&\simeq&-\frac{e^2F}{\pi m^*}
\sum_{p=1}^{\infty}\frac{A_p}{2\pi p}
\sin\left (2\pi p\frac{F}{B}-2\pi\gamma p+ \phi_p \right ),
\ee
where $A_p$ and $\phi_p$ are respectively the amplitude and phase of
the imaginary damping factor $Z_p$ defined by
\bb
\label{Zp00}
Z_p=A_p\exp(i\phi_p)=\frac{\pi}{2}\int_{-\beta\mu/(2\pi)}^{+\infty}
\frac{1+\tau x/\sigma+\omega x^2/\sigma}{
1+2\omega x/\tau }
\frac{\exp\Big [2i\pi px(\tau+\omega x)\Big ]}{\cosh^2(\pi x)}\d x,
\ee
and which involves the dimensionless parameters
\bb\label{param}
\sigma=\frac{F}{B},\;\;\tau=\frac{2\pi}{\beta\hbar\om},\;\;\omega=\frac{
\pi\tau\tk } { \beta}.
\ee
Hence, the temperature dependence is given both by $\tau\propto T$ and
$\omega\propto T^2$ while $\omega$, which is proportional to $S''(E)$,  includes
the band curvature contribution.
The integral can be computed by extending the lower bound to $-\infty$ at low
temperature and by neglecting the factors in front of the exponential that are 
proportional to parameter $\sigma^{-1}$ which is generally small compared to 
unity for magnetic fields strength currently available, except for very small 
orbits area. Besides, the
denominator $(1+2\pi x\tk/\beta)$ is strictly positive as long as $\tk
\mu<1$. In the peculiar case where $\tk=1/\mu$, which is relevant for Dirac 
fermions, as developed below, we can show that
$\omega=\tau^2/4\sigma$ and that the overall factor reduces actually to
$1+\tau x/(2\sigma)$ with no divergence around the lower bound of integration. Following the LK theory and using the
residue theorem, we can consider an integration path that goes
over the upper complex plane when $x$ is complex. Indeed, the singularities of
the $\cosh(\pi x)^{-2}$ function give the main contributions to the 
integral since they are located on the positive imaginary plane
$x_n=i(n+\ff)$. The oscillation
amplitude can be written as an infinite summation over the $x_n$'s
\bb
\label{Zp0}                               
Z_p=\sum_{n=0}^{\infty}\left \{
2\pi p\tau+4i\pi p\omega\Big (n+\ff
\Big )
\right \}
\exp\Big [
-2\pi p\tau\Big (n+\ff\Big )-2i\pi p\omega
\Big (n+\ff\Big )^2
\Big ].
\ee
For $\tk=0$ (or $\omega=0$), the summation can be performed and we obtain the
well known LK thermal reduction factor $Z_p=R_p=p\pi\tau/\sinh(p\pi\tau)$ 
\cite{Sh84}, to which
the Dingle factor and, eventually, magnetic breakdown probabilities are added. 
\eref{Zp0} can be expanded up to the
first
order in $\omega\propto\tk$, the sum over $n$ performed, and the result
re-exponentiated, which allows us to rewrite \eref{Zp0} as
\bb\nn\label{Zp}                                            
Z_p \simeq
\frac{p\pi\tau}{\sinh(p\pi\tau)}
\exp\left [ i \frac{\omega}{2\tau} \varphi\Big (\pi
p\tau
\Big ) \right ]=
R_p\exp\left [ i \frac{\pi}{2}\tk k_BT \varphi\Big ( 2\pi^2
p \frac{k_BT}{\hbar\om}
\Big ) \right ],\\
\varphi(x)=\Big [ \sinh(2x)-x-x\cosh^2(x) \Big ]/\sinh^2(x).
\ee
We can identify the amplitude $A_p$ with $R_p$ at this order. As a 
consequence, while
the oscillation amplitude, which is accounted for by the real part of $Z_p$, is 
unaffected by deviations from parabolicity, an extra phase is present in 
addition to the $\gamma$ constant in \eref{mosc}, and which is given by the 
complex argument. The expression of this additional phase is given 
by~\eref{phi} in the introduction, and is field- and temperature-dependent and 
proportional to the curvature factor $\kappa$.
It is worthwhile to notice that $\varphi(x)$ changes its sign
at $x = 1.606115$. Besides, when $x$ is large, i.e. B/T goes
to zero, $\varphi(x) \simeq 2-x$. Nevertheless $x$ is much larger than 2 in this range, hence $\varphi(x)\simeq -x$. Therefore,
the phase factor is given by
\bb
\label{phi_lB}
\phi_p \simeq -p\frac{2\pi^4k_B^2\hbar}{e}S''(E)\frac{T^2}{B}.
\ee
In the opposite case, for very large B/T ratio or very small effective mass,
$\varphi(x) \simeq x/3$, and the
Onsager phase can be approximated by
\bb
\label{phi_hB}
\phi_p \simeq p\frac{2\pi^4k_B^2\hbar}{3e}S''(E)\frac{T^2}{B}.
\ee
The above asymptotic expressions~\eref{phi_lB} and~\eref{phi_hB} indicate
a square temperature dependence and an inverse field dependence of $\phi_p$
both in the high an low field ranges. Besides, within the tight binding model,
$S''(\mu) = \Ar m^{*2}/(8 \pi \hbar^4)$, indicating
largest effect for large unit cell and effective mass.
These two features are achieved in organic metals. As an example, unit cell area
in the conducting plane as large as 108 ${\rm \AA}^2$\cite{Ur88,Ju89}  and
effective masses in the range $m_{\alpha}$ = 3 to 
3.51~\cite{He91,Me95,Ha96,Uj97,St99} have been reported for the $\alpha$ orbit 
of
$\kappa$-(ET)$_2$[Cu(NCS)]$_2$ (where ET stands for the
bis-ethelyne-dithio-tetrathia-fulvalene molecule).
$\kappa$ can be estimated from band structure
calculations and crystallographic data of
$\kappa$-(ET)$_2$[Cu(NCS)$_2$]\cite{Ur88,Ju89}, yielding e.g. 
$\kappa \sim$  5 $\times$ 10$^{58} {\rm m^{-2}J^{-2}}$ in the
$M-\Gamma$ direction, relevant to the $\alpha$ orbit.

\begin{figure}[ht]
\centering
\includegraphics[width=0.75\columnwidth,clip,angle=0]{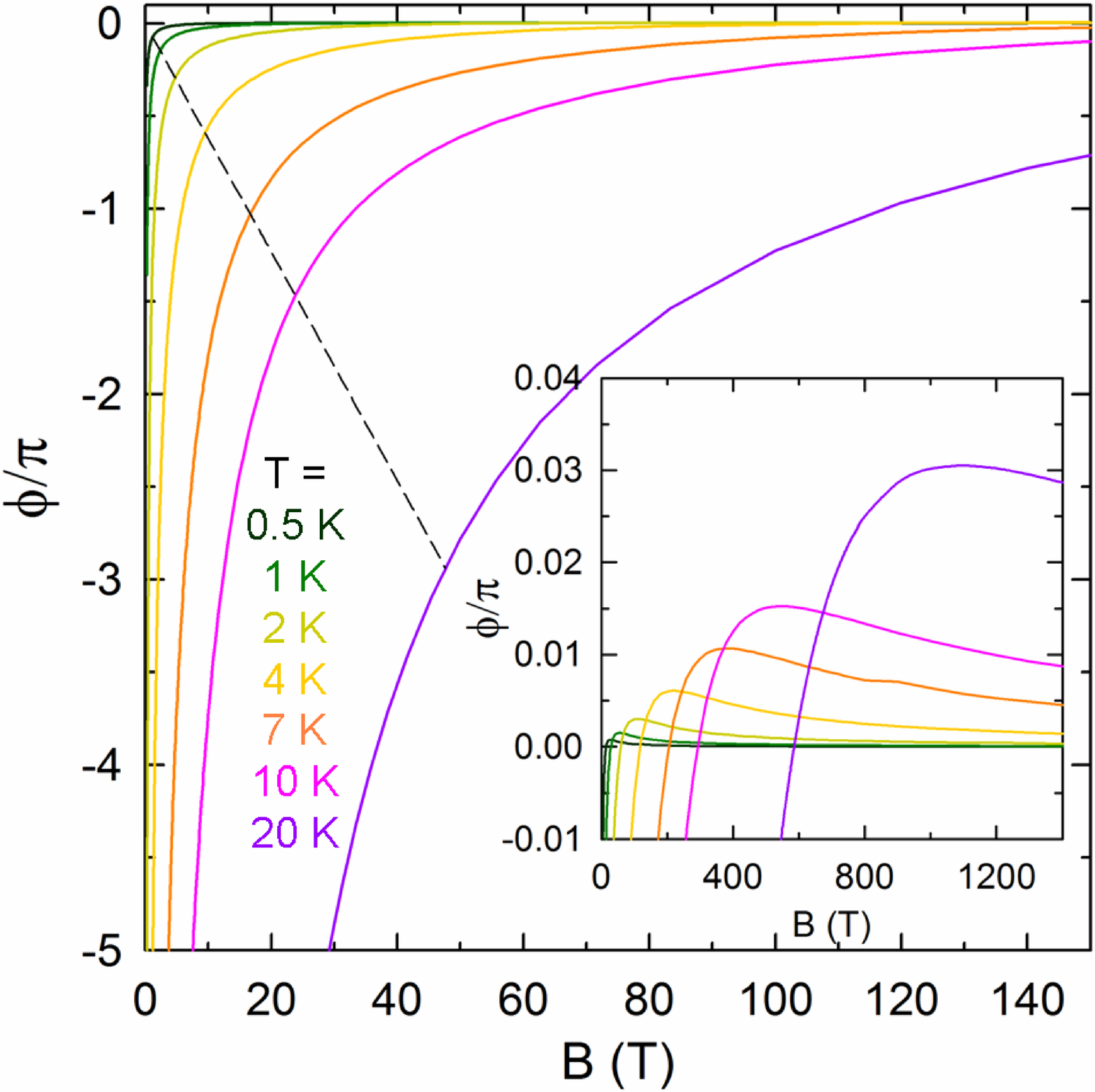}
\caption{Field dependence of the Onsager phase $\phi_1$ at various temperatures 
deduced from \eref{phi} for parameters relevant to the $\alpha$ orbit of the 
quasi-two dimensional organic metals $\kappa$-(ET)$_2$[Cu(NCS)$_2$] 
($\partial^2S/\partial E^2$ = 5 $\times$ 10$^{58}{\rm m^{-2}J^{-2}}$ and $m^*$ = 
3.2 $m_e$). The dashed line marks the field values such as $k_B T \simeq \hbar 
\omega_c$, which can be regarded as an estimation of the lower limit for the 
field above which oscillations can be observed.}
%
\label{fig:phi1}
\end{figure}
As displayed in \efig{fig:phi1}, we have plotted the Onsager phase $\phi_1$ 
versus field for several temperature values. The phase is 
very small at high field. Oppositely, large effects can be observed at low field
and high temperature. The dashed line determines the threshold limit when the 
Landau level gap is equal to the thermal fluctuations, 
$\hbar\omega_c\simeq k_BT$, and above which the oscillations can be observed 
at higher fields. Additionally, there is another limit based on the Dingle 
temperature $T_D$, which also imposes a minimum threshold for 
the field above which the oscillations can be seen. If we take $T_D=0.5$K (the 
best temperature for $\kappa$-CuNCS~\cite{bangura:2007}) 
one obtains a threshold field equal to 7.5 T when $\omega_c \tau=1$, and where 
$\tau=\hbar/k_BT_D$ is the scattering time. One however observes oscillations 
below this threshold limit in some compounds, when for example $\omega_c\tau$ 
is equal to 0.28~\cite{wosnitza:2003}.
To our knowledge, quantum oscillations in organic metals have been up to now 
observed above
several teslas at liquid helium temperatures. Therefore,
the studied effect could only be observed provided clean compounds with very small
scattering rate are synthesized. In this low field range, where $\phi_1/2\pi$
varies by several units, $\phi_p$ can
be rewritten as $\phi_p$ = $-2\pi p \delta F / B$ where $\delta F$ is given by:
%
%
\bb
\freq=\frac{2\pi^3}{\hbar e}S''(\mu)(k_BT)^2.
\ee
As a typical example, the tight binding model yields $$\freq=(k_BT)^2(\pi^2/24)m^{*2}\Ar/(e\hbar^3).$$ For $m^* = m_e$ and $\Ar=1{\rm \AA}^2$, i.e. for
parameters close to those of elemental metals, $\freq=3.462\times 10^{-6}T^2$ 
which remains negligibly small, even at high temperature. For the above 
considered
organic metal, temperature as high as few tens of a Kelvin is nonetheless necessary to get a frequency shift of few T. Namely, $\delta F$ value as low as  1.5 T should be observed at 20 K which could hardly be detected owing to the oscillation frequency value $F_{\alpha}$ = 600 T
yielding $\delta F/F$ = 0.25 $\%$.
\section{Dirac Fermions} 

\medskip

In this section, we consider the case of Dirac fermions such as observed in monolayer graphene which has been intensively studied (for a
review, see e.g. Refs.~\cite{Go11,Or13}). The data are in agreement with a
linear band dispersion, the curvature parameter being given by
$S''(E) = 2\pi/(\hbar v_F)^2$ with a Fermi velocity $v_F =
10^6{\rm ms^{-1}}$~\cite{Zh05,No05,Sa06}.
Shubnikov-de Haas (SdH) oscillations with effective mass in the range
$7\times 10^{-3}$ $m_e$ to 0.04 $m_e$, depending on the carrier concentration driven
by bias voltage, have
been observed at temperatures either up to 50 K  \cite{Zh05} or above 100 K
\cite{No05}, allowing to expect detection of larger values of the phase $\phi_1$.

Contrary to the parabolic case, the Landau level energies $E_n$ are not
linear with index $n$ but are given by $E_n=E_1\sqrt{n}$, with
$E_1^2=2e\hbar v_F^2B$. Since $E_n'\propto n^{-1/2}$, $E_n'^2\propto n^{-1}$ 
and $E_n''\propto n^{-3/2}$, the two terms in the right hand side 
of~\eref{omega1} are diverging when $n$ goes to zero. Therefore, the 
intermediate step to compute
$\Omega_{osc}$ using two integrations by parts in \eref{omega1} cannot be
performed here, since these contributions  makes the integral in
\eref{omega1} divergent. Taking into account the band degeneracy $g_D=4$ in the 
case of graphene, we can write instead
\bb\nn
\frac{\Omega_{osc}}{A}&=&-2g_Db\,\Re
\sum_{p=1}^{\infty}
\int_0^{\infty}\frac{\exp(2i\pi
pn)}{1+\exp[\beta(E_n-\mu)]}\frac{E_n'\,\d n}{2i\pi p}
\\
&=&-2g_Db\,\Re \sum_{p=1}^{\infty}\frac{1}{2i\pi p}
\int_0^{\infty}
\frac{\exp(2i\pi pE^2/E_1^2)}{1+\exp[\beta(E-\mu)]}\,\d E.
\ee
An integration by parts is possible if we integrate $\exp(2i\pi pE^2/E_1^2)$
using the error function in the complex plane
\bb
\frac{\Omega_{osc}}{A}=2g_Db\,\Re
\sum_{p=1}^{\infty}\frac{\beta
E_1}{4(-2i\pi p)^{3/2}}\frac{\sqrt{\pi}}{2}
\int_0^{\infty}
\frac{\erf(\sqrt{-2i\pi p}E/E_1)}{\cosh^2[\beta(E-\mu)/2]}\,\d E.
\ee
We can set as before $x=\beta(E-\mu)/2\pi$ and obtain
\bb\label{OmegaGraph}
\frac{\Omega_{osc}}{A}=2g_Db\,\Re
\sum_{p=1}^{\infty}
\frac{E_1}{2(-2i\pi p)^{3/2}}\frac{\pi^{3/2}}{2}
\int_{-\beta\mu/(2\pi)}^{\infty}
\frac{\erf\Big (\sqrt{-2i\pi p}\{\mu+2\pi x/\beta\}/E_1\Big )}{\cosh^2(\pi
x)}\,\d x.
\ee
The dominant part of the magnetization is
obtained by differentiation of the erf function with respect to minus $B$ in the
previous expression, since the dominant oscillating term comes from this
function
\bb\nn
m_{osc}&=&-g_D\frac{e\mu}{h}\Re \sum_{p\ge 1}\frac{\exp(2i\pi
pF/B)}{2i\pi
p}Z_p,
\\ \label{ZpGraph}
Z_p&=&\frac{\pi}{2}\int_{-2\sigma/\tau}^{+\infty}\left
(1+\frac{\tau x}{2\sigma}\right )\frac{\exp(2i\pi p[\tau
x+\tau^2x^2/(4\sigma)])}{\cosh^2(\pi x)}\d x.
\ee
$Z_p$ is normalized such that $Z_p=1$ at
zero temperature, i.e. $\tau=0$, using
the integral value $\pi\int_0^{\infty}\cosh(\pi x)^{-2}\d x=1$.
In this expression, the dominant frequency $F$ is defined by 
$\mu^2/E_1^2=F/B=\sigma$, and the damping factor $\tau=4\pi\mu/(\beta
E_1^2)$ which is temperature-
and field-dependent. These parameters are expressed as

\bb                                                          
F=\frac{\mu^2}{2e\hbar v_F^2},\;\tau=\frac{2\pi\mu}{\beta
e\hbar
v_F^2B},\;\;\sigma=\frac{F}{B},\;\;\frac{\tau}{\sigma}=\frac{4\pi}{\beta\mu}.
\ee

In reference~\cite{shap04}, the authors obtained the expression for the
zero-temperature thermodynamic potential, given here by ~\eref{OmegaGraph}, in terms of a series
involving Bernoulli polynomials, up to the fourth power in magnetic field. This
is very similar to the expansion obtained for the general case from Eqs.~(\ref{Sa}) and (\ref{omega2}) in the
sense that curvature induces an expansion over the cyclotronic frequency. As
in~\eref{omega2}, Bernoulli polynomials $B_n(x)$ of order $n$ are
periodic functions of $x$ with harmonics $p$ that decay like $p^{-n}$.
In~\eref{omega2} this corresponds to an expansion up to $n=3$ that includes the
temperature dependence. In the case of Dirac fermions, this expansion is hidden in the
factor $Z_p$ whose dependence on the harmonics index $p$ can be explicitly
determined by expanding the $x^2$ argument in the exponential of~\eref{ZpGraph} as
a series. The remaining integrals can in principle be performed in the complex
plane yielding a power series in terms of $1/p$.
In the limit where $\sigma\gg\tau$, we
have $Z_p=\pi p\tau/\sinh(\pi p\tau)=R_p(\pi\tau)$. We may then
identify
$\pi\tau$ with $2\pi^2/(\beta\hbar\omega_c)$ as in \eref{param}, by
defining the cyclotronic frequency $\omega_c=eB/m^*$ and effective
relativistic mass $m^*=\mu/v_F^2$. Otherwise, the amplitude
factor $Z_p=A_p\exp(i\phi_p)$ is complex and possesses a modulus
$A_p$ and non-zero argument $\phi_p$ contributing to the global
Onsager phase of the oscillations.

We may apply the residue theorem to the previous integral, in the limit where
the lower bound $-2\sigma/\tau$ is large, or typically $T<m^*v_F^2/k_B$.
The singularities of the function $\cosh(x)^{-2}$ are still located at
$x_n=i(n+\ff)$ and we obtain
\bb\label{grapheneZp}
Z_p\simeq
\sum_{n\ge 0}
\left \{
2p\pi\tau\left [1+\frac{i\tau(n+\ff)}{2\sigma} \right ]^2+
\frac{\tau}{2i\sigma}
\right \}
\exp\Big [-2\pi p\tau (n+\ff)-2i\pi p\frac{\tau^2}{4\sigma}(n+\ff)^2\Big ].
\ee

\begin{figure}[ht]
\centering
\includegraphics[width=0.75\columnwidth, clip,angle=0]{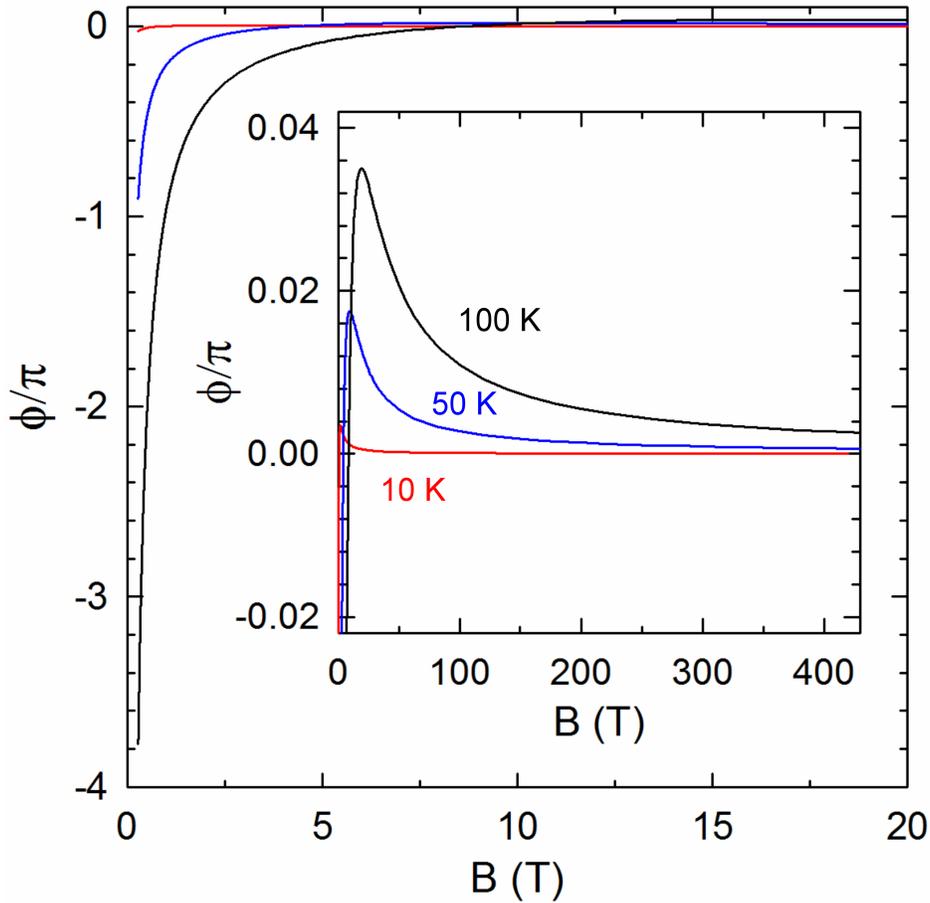}
\caption{Field dependence of the Onsager phase $\phi_1$ for Dirac fermions 
derived from~\eref{phiGraph} with
Fermi velocity $v_F$ = 10$^6{\rm ms^{-1}}$ and effective
mass $m^*$ = 0.02 at T = 10 K, 50 K and 100 K.}
\label{fig:phi1_Dirac}
\end{figure}

This expression can be compared to~\eref{Zp0} provided substitutions
$\tk=1/\mu$ and $\omega=\tau^2/4\sigma$ are made. Except for the argument of the
exponential terms, \eref{grapheneZp} is different from the general
case, yielding therefore a modified form of the Onsager phase $\phi_p$ and 
function $\varphi(x)$ defined in \eref{Zp}, given by~\eref{phiGraph}.
%
%
In this case, the series expansion valid at low $m^*T/B$ ratio gives $\varphi(x)\simeq x$ instead of
$x/3$, and the phase $\phi_p$ is inversely proportional to the magnetic field.
The other limit as $m^*T/B$ is large, is as previously negative, although
$\varphi(x)\simeq 4-x$ instead of $2-x$. As in the previous case, $x$ is much 
larger than 4 in this range, hence $\varphi(x)\simeq -x$. Therefore,
the phase factor is given by
\bb
\label{eq:phi_p_approx}
\phi_p=\pm\frac{2\pi^3p}{\hbar eB}\left(\frac{k_BT}{v_F}\right)^2,
\ee
where the sign plus and minus corresponds to the high and low field limit, 
respectively. According to~\eref{eq:phi_p_approx}, $\phi_p$ depends on the 
Fermi velocity, instead of the effective mass and band curvature, at both high 
and low field.  Field dependence of $\phi_1$ is displayed in 
Fig.~\ref{fig:phi1_Dirac} for experimental parameters relevant to
graphene \cite{Zh05,No05}. As reported in the preceding section, negligibly
small values are obtained in the high field range. In the low field range, the
same behaviour as for organic metals is obtained. Namely, a significant drop down
of $\phi_1$ at low field and high temperature. Nevertheless, the frequency
shift, which can be written  $\freq/F=2\pi^2\left
(\frac{k_BT}{m^*v_F^2}\right )^2$, remains small (about 8 \% at 100 K in the
example of Fig.~\ref{fig:phi1_Dirac}) although higher than in the case of the
organic metal considered in the preceding section.

\section{Summary and conclusion}

We have evidenced that a non-parabolicity of the band dispersion versus momentum 
yield a field-dependent shift of the Onsager phase of quantum oscillations, i.e. 
a frequency shift. The analytical expressions of this phase shift depend on 
the nature of the dispersion band, either mainly parabolic with corrective 
terms to the
parabolicity (organic conductors) or linear (Dirac fermions). It is demonstrated that,
in both cases, the effect is amplified at low magnetic field, high temperature
and either large effective mass and unit cell area (organic conductors) or small
Fermi velocity (Dirac fermions). However, the observed effect is small within
the experimental conditions where quantum oscillations are observed.
Largest effect are nevertheless observed for Dirac fermions. Indeed, the 
deviation from parabolicity of their band dispersion is the most significant. 
Since the phase shift is expected to be the largest for small Fermi velocity,  
Dirac fermions in organic conductors, such as observed in 
$\alpha$-(BEDT-TTF)$_2$I$_3$ \cite{Ka06,Pi13}, with large unit cell, hence small 
Fermi velocity appear as promising candidates.
It is expected that samples should be clean enough to observe the large phase 
shift deviations at low fields. This typically corresponds to 
the integer quantum Hall regime, which requires a proper treatment of the 
disorder effects but are not that relevant for the magnetization oscillations, 
in contrast to transport coefficients. The temperature-dependent phase shifts 
are not expected to be fundamentally
affected.  Indeed, the principal modification for the magnetization
oscillations in the quantum Hall regime will take place at the level of
the disorder Dingle factor (associated with a
Lorentzian distribution of the Landau bands) and may be no more valid
in low-dimensions, and can be typically replaced by a Gaussian reduction
factor (with a Gaussian dependence on the harmonics $p$), which reflects a 
Gaussian shape of the Landau bands as often reported experimentally for the 
density of states in the integral quantum Hall regime~\cite{Fogler:2002}.
Also, it may be important to consider theoretical Fermi surfaces with 
singular behavior which is affected by large curvature and therefore may 
display important phase 
shifts.

\bibliography{curvature}

\end{document}